\documentclass[aps,prl,showkeys,superscriptaddress,10pt,nobibnotes,twocolumn]{revtex4-1} 

\usepackage{amsmath,bm}
\usepackage{graphicx}
\usepackage[dvipsnames]{xcolor}

\newcommand{\lw}{\linewidth}
\graphicspath{{./fig/}}
\begin{document}
\title{Soft magnon contributions to dielectric constant in spiral magnets with domain walls}

\author{Francesco Foggetti}
\affiliation{Quantum Materials Theory, Italian Institute of Technology, Via Morego 30, 16163 Genova, Italy}
\affiliation{Department of Physics, University of Genova, Via Dodecaneso, 33, 16146 Genova GE}
\author{Sergey Artyukhin}
\affiliation{Quantum Materials Theory, Italian Institute of Technology, Via Morego 30, 16163 Genova, Italy}
\begin{abstract}
Competing magnetic exchange interactions often result in noncollinear magnetic states, such as spin spirals, which break the inversion symmetry and induce ferroelectric polarization \cite{Cheong2007}. The resulting strong interactions between magnetic and dielectric degrees of freedom, lead to a technologically important possibility to control magnetic order by electric fields \cite{KimuraNat2003,Leo2015} and to electromagnons, magnetic excitations that can be excited by an electric dipole of the electromagnetic field \cite{Pimenov2006}. Here we study the effects of chiral domain walls on magnetoelectric properties of spiral magnets. We use a quasi-1D model Hamiltonian with competing Heisenberg exchange interactions, leading to a spin spiral, and Dzyaloshinskii-Moriya interactions, that couple spins and electric dipoles and mix magnon and phonon excitations. The results suggest that low frequency dielectric anomalies in spiral magnets, such as TbMnO$_3$ and MnWO$_4$, may originate from hybrid magnon-polar phonon excitations associated with domain walls.
\end{abstract}

\maketitle

\textit{ Introduction} -- Magnetic frustration usually provides novel and interesting magnetic and dielectric properties \cite{Cheong2007} as the non trivial structure the spins arrange themselves into becomes the playground for excitations of magnetic and, if magnetic and electric degrees of freedom are related, electric kind \cite{ValdesAguilar2009}. In the ordered phase, an inhomogeneous spin texture can be observed, with multiple magnetic domains separated by domain walls (DWs), with non-trivial domain patterns, e.g. vortices \cite{Li2012,Schoenherr2018,Nattermann2018}. DW are no mere transition regions between different domains but dynamical objects that can interact with excitations, move and alter the properties of a material.

The contemporary presence of magnetic frustration, domain walls and magnetoelectric (ME) coupling in a material or in a class of materials makes for an ideal candidate of a detailed study as the possible applications of such a system include sensing, information technology, next-generation spintronic devices \cite{Bader2010}.
  
In helimagnets with competing nearest- and next-nearest neighbor exchange interactions, such as TbMnO$_3$, MnWO$_4$, CuO, magnetic frustration forces spins to assume a spiral configuration \cite{Kimura2003,Nattermann2018}. ME coupling via Dzyaloshinskii-Moriya interaction then induces opposite ferroelectric polarizations in the helical domains with opposite spin rotation sense (opposite chirality). Here ``chiral'' is understood as not identical to its mirror image, and, in the context of cycloidal spin spiral also implies an associated sign of a ferroelectric polarization \cite{Katsura2005,Mostovoy2006,Sergienko2006}. Soft DW-localized magnetic excitations, related to DW motion, that are electrically active, may be expected to affect the dielectric properties \cite{Kagawa2009}.

The presence of a DW is of a fundamental importance because it alters the spectrum of spin excitations (magnons) and polar modes (polar phonons) by introducing DW localized modes, representing DW motion or deformations that can interact with the bulk modes \cite{Brierley2014}. Electric and magnetic excitations are intertwinned due to ME coupling, and some of them acquire a uniform polarization component, giving rise to electromagnons \cite{Pimenov2006}. The contribution of the DW-localized modes to the macroscopic properties is proportional to the volume fraction DWs occupy, which is usually small. It is enhanced in natural multidomain states or in engineered domain wall arrays \cite{Hlinka2017}. On the other hand the contribution of these soft modes to the low-frequency dielectric response is proportional to $1/\omega^2$ which may make it noticeable. Indeed, a dielectric relaxation has been observed around 10 MHz, near the domain wall-generating spiral flop transition~\cite{Kagawa2009,Kagawa2011}. The dependence of the intensity on poling suggests that the relaxation is related to the presence of DWs. 
The papers by Schrettle \cite{Schrettle2009} and Schiebl \cite{Schiebl2015} also point the difference in relaxation and strong magnetocapacitive behavior between spiral and A-type magnets. In addition to the MHz-range mode, Schiebl~\cite{Schiebl2015} also describes the relaxation 6 orders of magnitude lower in frequencies compared to Refs.~\cite{Kagawa2009,Kagawa2011}. Schrettle~\cite{Schrettle2009} measurements are not directly comparable since due to a different geometry ($E\|c$).

In this Letter we study dielectric properties of non-chiral helimagnets (having no preferential sense of spin rotation) due to the presence of DWs. Interacting spins and polar distortions are modeled by means of a model Hamiltonian. We study the spectrum of excitations of a uniform spiral ground state and a system with chiral domain walls, and discover domain wall localized electromagnons and the connection between these modes and the magnetoelectric response. The results emphasize the importance of ME coupling and the features of magnon and phonon spectra in helical magnets with DWs, and help interpret the experimental data on microwave dielectric loss in these systems.

\begin{figure}[t]
    \centering
    \includegraphics[width=.9\lw]{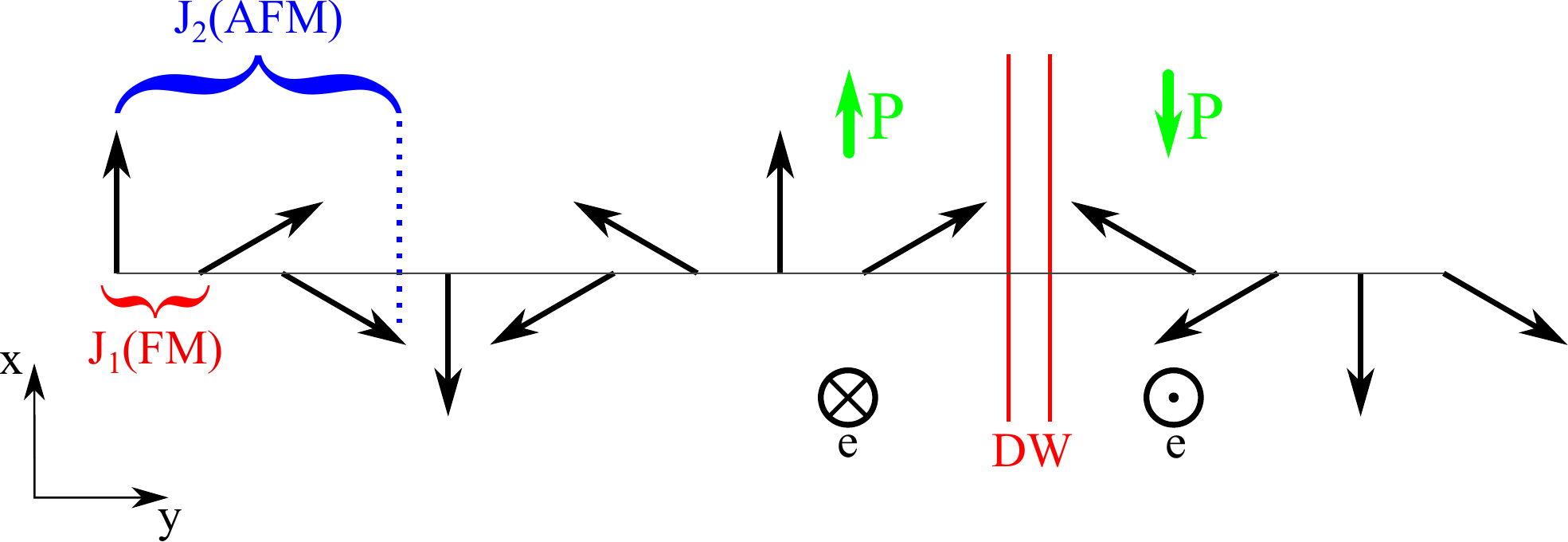}
    \caption{\label{spir}(Colors online) Schematics of a 1D spiral system. The competing FM and AFM interactions on the spins generate a frustrated spiral structure while the DM term associates an electric polarization (green) to the spiral. As the spiral changes its chirality two chiral domains and a DW are identified, polarization is opposite in the two domains.}
\end{figure}

\textit{The Model} --
Here the effects of the presence of a DW are studied in a geometry where two different chiral domains (magnetic domains with opposite spin rotation axes) meet at a planar DW, perpendicular to the $y$ axis, as shown in Fig.~\ref{spir}. 

\begin{figure*}[t]
    \centering
    \includegraphics[width=\lw]{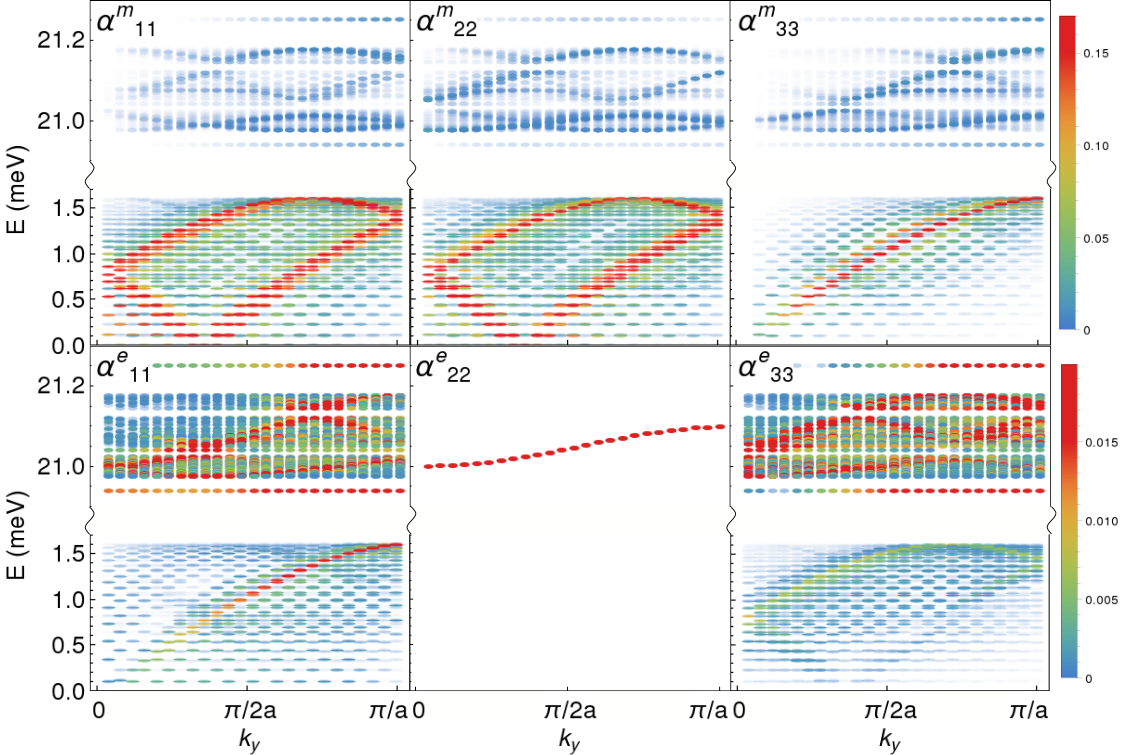}
    \caption{\label{sus} (Colors online) Absolute values of the diagonal components of magnetic (top row) and electric (bottom row) susceptibilities $|\alpha^{m,e}_{ii}(k_y,\omega)|$ for a spiral system with a domain wall. Colorscale encodes the absolute value of the susceptibility, and a low saturation level is chosen to highlight the low-energy modes. The commonalities between electric and magnetic susceptibility indicate the mixing of electric and magnetic excitations due to DM interaction. Off-diagonal components of $\alpha^e$ are zero. There are no excitations inside the cropped energy interval between 1.7 and 20~meV.}
\end{figure*}

Although non trivial spin structures appear in both bulk crystals and lower dimensional systems \cite{Li2012,ValdesAguilar2009,Borisov2005,Schoenherr2018} here we employ a minimal quasi-1D model to capture essential physics of helical magnets, describe domain wall dynamics and study the contributions of electric and magnetic excitations to the dielectric character of the class of materials that exhibit this peculiar spin ordering.
We use the following Hamiltonian,
\begin{equation}
\begin{split}
    H&=\sum_i J_1\mathbf{S}_i\cdot \mathbf{S}_{i+1}+J_2\mathbf{S}_i\cdot \mathbf{S}_{i+2}+K_h(S_i^z)^2\\  &+\dfrac{1}{2m}\pi_i^2-V_{uc}\mathbf{P}_i\cdot \mathbf{E}+\frac{c}{2}(\mathbf{\delta r}_{i+1}-\mathbf{\delta r}_i)^2 +\dfrac{m\bar\omega^2}{2}\delta r_i^2\\  &+\alpha_{DM}(\mathbf{r}_{12}\times \mathbf{\delta r}_i)\cdot(\mathbf{S}_i\times \mathbf{S}_{i+1}),
    \label{ham}
\end{split}
\end{equation}
where $S_i$ is the classical spin ($|S|=1$) at site $i$, terms with $J_1$ and $J_2$ represent competing Heisenberg interactions, ferromagnetic (FM) between nearest neighbour spins (NN) and antiferromagnetic (AFM) between next nearest neighbours (NNN), respectively; the term with $K_h$ is a hard $z-$axis anisotropy that forces the spins into a cycloidal plane ($xy$ plane) as observed e.g. for TbMnO$_3$ \cite{Senff2007}. In orthorhombic systems an easy axis anisotropy within the spiral plane, neglected here,  may lead to a lock-in transition to a commensurate state, resulting in immobile chiral DWs, which may explain the results in Ref.~\cite{Lottermoser2004}. $\mathbf{P}_i=Z\mathbf{\delta r}_i$ is the polarization due to the polar mode (electric dipole per unit cell volume $V_{uc}$), representing antiphase shift of oxygen and Mn ions in the unit cell $i$, $\pi_i$ is the associated canonical momentum; $c$ the polar mode stiffness, $m$ -- the mode effective mass divided by the square of the mode effective charge $Z$, $\mathrm{E}$ is the external electric field, $\bar \omega$ is the frequency of the polar optical mode \cite{Sergienko2006}.
The term with $\alpha_{DM}$, Dzyaloshinskii-Moriya (DM) interaction \cite{Dzyaloshinskii60,Moriya60,Dzyaloshinskii64}, couples the oxygen shifts and spins and is responsible for the magnetoelectric effect in the spiral state: a spin excitation will alter the ionic displacements thus generating a response in the polarization and vice versa. $\mathbf{r}_{12}=(\mathbf{r}_{i+1}-\mathbf{r}_i)\|\hat y$ represents the vector connecting the neighboring magnetic ions. The values of the parameters used in the calculations are $J_1=0.3$~meV, $J_2=0.1$~meV, $V_{uc}=230$~\AA$^2$ (similar to those in TbMnO$_3$ \cite{Senff2008,Senff2007}), $\bar\omega=21$~meV$=5$~THz \cite{Schmidt2009}, $K_h=0.1$~meV, $m\approx 3m_O/V_{uc}$ with $m_O$ being an oxygen mass, $\alpha_{DM}=1$~meV/\AA$^2$, $c=0.3$~meV/\AA$^2$.
Effects due to symmetric exchange striction may also be important~\cite{Aguilar2009} and will be the topic of the following work.

\textit{Spiral solution} -- For an AFM NNN interaction competing with the NN exchange, the equilibrium spin configuration cannot be of FM or AFM type. The interplay between these exchange interactions generates magnetic frustration thus resulting in a spiral configuration. As the cycloidal spiral (with the rotation vector $\mathbf{e}$ perpendicular to the propagation vector $\mathbf{k}$) breaks the inversion symmetry, the spin canting induces ionic displacements through the DM interaction. The DM interaction interlaces the spin spiral with the electric polarization. In this way ME coupling is achieved.

In this work we analyze a spiral configuration in which the chirality, the spin rotation vector $\mathbf{e}_i=[\mathbf{S}_i\times \mathbf{S}_{i+1}]$ of the spiral $\mathbf{e}$, is reversed across the DW, hence defining two chiral domains separated by a chiral domain wall (DW) as seen in Fig.~\ref{spir}. 


\textit{Results} -- To characterize the excitation spectrum in the presence of a chiral DW we expand the Hamiltonian Eq.~(\ref{ham}) around a local minimum, corresponding to two adjacent chiral domains, separated by two DWs, in the quasi-1D geometry with open boundary conditions, cf.~Fig.~\ref{spir}. One DW is at the center, indicated by red vertical lines in Fig.~\ref{spir}, and one at the boundary. We linearize the equations of motion in the spirit of linear spin wave theory (see Supplementary for the details \footnote{Link to Supplementary to be inserted}), and the resulting eigenproblem gives coupled magnon-phonon modes. Given the dimension of the chain and a large resulting number of modes (50 ions in the chain and 400 modes), the spectra are rather complex, as seen in Fig.~\ref{sus}, but the effects of the domain wall appear to be non-trivial.

The modes at the highest energy, $E\sim 21$~meV originate from polar phonons, corresponding to antiphase motions of magnetic ions and oxygens, the energy scale for which is $\hbar\bar\omega$. The modes below 2~meV mostly have magnon character. The V-shaped dispersion originating from $q=Q_{s},\omega=0$, the Bragg peak of the spiral, corresponds to a phason mode. DM interaction mixes polar phonons and magnons, therefore common features appear in electric and magnetic susceptibility. The only exception is $\alpha^{e}_{22}$, the figure shows only a thin red line because the bond vector $\mathbf{r}_{12}\|\hat y$ so the DM interaction does not involve $P_y$. The polar phonons, polarized along $x$ and $z$ broaden due to mixing with magnons. The cross product structure of the DM term is responsible for the mixing of the magnetic and electric susceptibility: the electric susceptibility $\alpha_{11}^{e}$ ($\alpha^{e}_{33}$) shows at low energy features from the magnetic susceptibility $\alpha_{33}^m$ ($\alpha_{11}^m$) as well as the magnetic one shows features from the electric one at high energy. The upper and lower subbands around 21~meV correspond to the polar phonons with an admixture of the optical and acoustic magnons, respectively. Separated by a gap, below 2~meV are the bulk magnon branches with a large dispersion (with an admixture of polar phonons), with spin oscillations perpendicular to the easy plane at 1-1.5~meV and below 0.5~meV. The intensity between the bright red bands is introduced by the presence of the DWs, as seen from the comparison with the magnon spectrum, calculated without the DWs, shown in Fig.~\ref{supME} in the Supplementary.

\begin{figure*}[t]
\includegraphics[width=\linewidth]{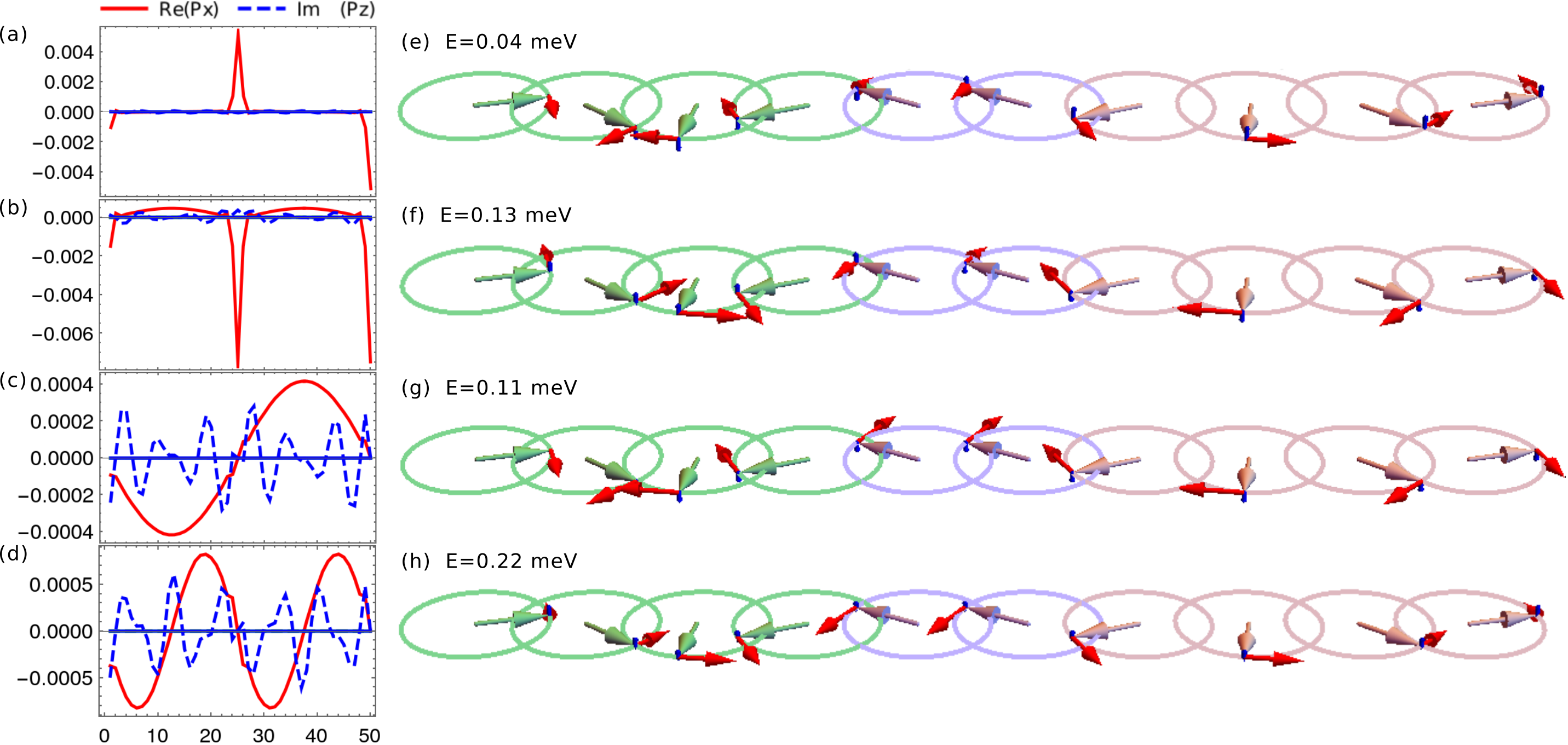}
\caption{\label{fig:modes}Soft hybridized magnon-phonon modes at the chiral domain wall. (a-d) Electric polarization profiles of the lowest-energy modes. The components that are not shown are zero. $P_y$ phonons do not mix with these modes. (e-h) Spin deviations for the same modes. The labels indicate the mode energies and the relative contributions of the polar phonons of different polarizations. The remaining contribution is from magnons. The Mn spins are indicated with the lines inside the circles. Green spins in green belong to clockwise chiral domain; orange spins-to counterclockwise domain, while the spins at the DW are in violet. The in-plane spin deviation corresponding to the mode is shown by red lines, tangent to the circles, and (much smaller) perpendicular components are in blue.
}
\end{figure*}

At the lowest frequencies are the dispersionless signatures of DW-localized modes, illustrated in Fig.~\ref{fig:modes}.
The bulk magnon modes, whose polarization texture is even with respect to the wall, shown in Fig.~\ref{fig:modes}(a-b), mix with the DW-localized magnons. The mixing is controlled by the strength of the DM interaction. These hybridized bulk-DW modes are spread through the lower part of the spectrum seen in Fig.~\ref{sus}. The magnons with an odd polarization texture (with a polarization node at the DW) are delocalized and unaffected by the presence of the wall, as seen in Fig.~\ref{fig:modes}(c-d). The lowest energy modes, seen in Fig.~\ref{fig:modes}(a-b), are associated with the sliding of the DW, and since the chiral wall is also ferroelectric, these modes change the electric polarization and are electromagnons. The one at the lowest energy, shown in Fig.~\ref{fig:modes}(a), corresponds to both walls shifting in the same direction (``acoustic mode''), while the other mode, in Fig.~\ref{fig:modes}(b) corresponds to opposite shifts of the walls, and hence the overall change of the polarization.  
For widely separated walls the splitting between these modes, induced by wall-wall interactions, must be small. A splitting of 0.09~meV, observed in our simulation, is a finite system size effect. The central frequency, 0.1~meV, is overestimated due to a very narrow wall, owing to a strong hard axis anisotropy. In TbMnO$_3$ and other spiral magnets, the electromagnon is much softer, leading to a large contribution to the dielectric constant of a spiral state $\epsilon=1+4\pi\alpha^e$, related to the electric susceptibility $\alpha^e$. Indeed, the equation of motion for the mode with the amplitude $x$, mode effective charge $q$ and mass $m$ under an external electric field $E(r,t)=Ee^{-i\Omega t}$ is
\begin{equation}
    m(\ddot x +\bar\omega^2 x)=q E e^{-i\Omega t}.
\end{equation}
Substituting  $x(t)=\Tilde{x}e^{-i\Omega t}$ we find the response $\Tilde{x}$ at the driving frequency $\Omega$,
$\Tilde{x}=qE/m(\bar\omega^2-\omega^2),$
and therefore the electric susceptibility is
\begin{equation}
    \alpha^{e}(\omega)=\frac{q\Tilde{x}(\omega)}{E(\omega)}=\frac{q^2}{m(\bar\omega^2-\omega^2)}.
\end{equation}
The static electric susceptibility
is ${\alpha^{e}(\omega=0)=q^2/m\bar\omega^2}$, thus the response is dominated by low-frequency modes. However, the mode effective charge $q=2PV_{\mathrm{DW}}/(mV)$ of the DW sliding mode involves the ratio of the DW volume to the entire sample volume, $V_{\mathrm{DW}}/V$, and large DW densities are therefore necessary for this contribution to be maximized.
Interactions of bulk and DW-localized electromagnons may lead to fascinating phenomena as the DW could filter certain modes while letting others propagate \cite{Brierley2014,Royo2017}. As DWs are easily writable and malleable with moderate external fields, these DW-localized electromagnons may be easily exploited in realization of spin based devices.



\begin{figure*}[ht]
    \centering
    \includegraphics[width=.85\lw]{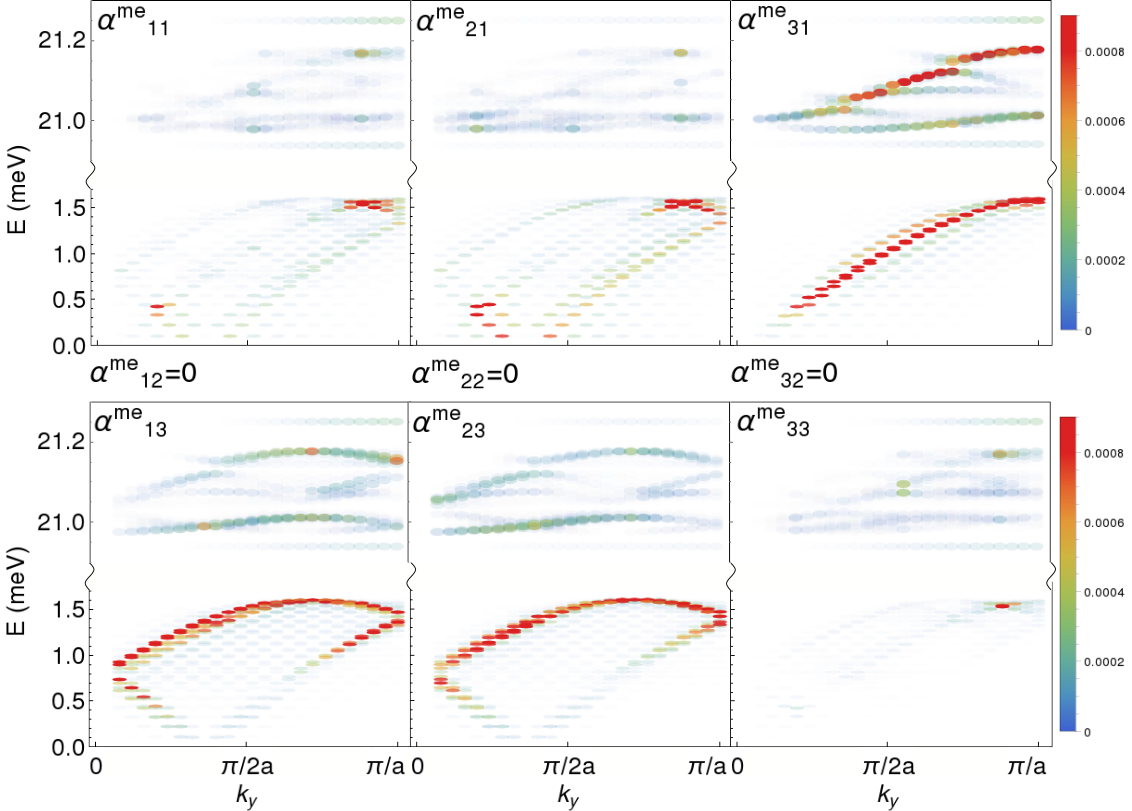}
    \caption{\label{fig:me} Magnetoelectric susceptibility $\alpha^{me}$ computed using Fourier component of M and P eigenvectors of the equations of motion. $\alpha^{me}_{i,2}=0$ due to the symmetry of DM terms.    There are no excitations inside the cropped energy interval between 1.7 and 20~meV.}
\end{figure*}

\textit{Magnetoelectric effect} -- The eigenmodes of the coupled spin-phonon system Eq.~\ref{ham} give access to ME susceptibilities $\alpha^{me}_{ij}(q,\omega)=\partial B_i/\partial E_j$ and $\alpha^{em}_{ij}(q,\omega)=\partial D_i/\partial H_j$ whose components are connected by Onsager relations $\alpha^{em}_{ij}(\omega,\vec H,\vec M)=-\alpha^{me}_{ji}(\omega,-\vec H, -\vec M)$. These definitions of $\alpha^{me}_{ij}$ and $\alpha^{em}_{ij}$ allow to obtain them from the calculated components of the phonon-magnon eigenvectors, cf. Fig.~\ref{fig:me}. 


The components $\alpha^{me}_{i,2}$ are zero since $P_y$ does not couple to spins in Eq.~\ref{ham} when the wave vector $\vec k\|\hat y$. Since the coupled magnon-phonon modes mediate the ME coupling, the branches in $\alpha^{me}_{ij}$ resemble the magnon bands in Fig.~\ref{fig:modes}.

Since the spiral breaks the translational symmetry, the generalized magnetoelectric tensor, dependent on two wave vectors $\alpha^{me}_{ij}(q_1,q_2,\omega)=\partial B_i(q_1)/\partial E_j(q_2)$ has a non-trivial structure. Particularly, replicas appear at $q_1=q_2\pm Q_s$, as seen in Fig.~S2. The presence of DWs further complicate this structure, since, due to sharpness of domain walls, they produce the potential in the magnon-phonon equations of motion, that scatters by wave vectors up to $k\sim 2\pi/\lambda_{DW}$. 

\textit{Conclusions} --
Magnetic and lattice excitations in a helical magnet are computed using a quasi-1D model in the presence of chiral domain walls. Domain wall-localized soft electromagnons are found and their contributions to dielectric properties and ME effect are studied. Frequency-dependent electric and magnetoelectric susceptibilities demonstrate composite magnetoelectric excitations in the broad frequency range. Results suggest that low energy DW-localized modes may dominate microwave dielectric response and may allow phonon and magnon filtering. These interesting dielectric and  magnetoelectric properties suggest the use of frustrated magnets as a materials platform for optical and spintronic devices.

\bibliography{biblio}

\widetext
\newpage

\begin{center}
\textbf{\large SUPPLEMENTARY INFORMATION}
\end{center}
\setcounter{table}{0}
\setcounter{figure}{0}
\renewcommand{\thetable}{S\arabic{table}}
\renewcommand\thefigure{S\arabic{figure}}
\vspace{0.3cm}

{\large \textbf{S1. Equation of Motion and susceptibility}}\\
In order to compute the susceptibility from Hamiltonian (\ref{ham}) we identify the ground stati in the spiral configuration with a DW in the center.  We define $\bm{\xi}$ so that  $\xi_i=(\theta_i,\phi_i,\bm{\pi}_i,\delta\bm{r}_i)$, and $\bm{\xi}_0$ will be the values of variables in the ground state. For small deviations $\Tilde{\bm{\xi}}$ it results $\bm{\xi}=\bm{\xi}_0+\Tilde{\bm{\xi}}$. We recompute the Hamiltonian up to the second order in the deviations $\Tilde{\bm{\xi}}$
\begin{equation}
    \mathcal{H}=\begin{pmatrix}
    \partial^2_{\phi\theta}&\partial^2_{\phi^2}&\partial^2_{\phi\pi}&\partial^2_{\phi\delta r}\\
    -\partial^2_{\theta^2}&-\partial^2_{\theta\phi}&-\partial^2_{\theta\pi}&-\partial^2_{\theta\delta r}\\
    -\partial^2_{\delta r\theta}&-\partial^2_{\delta r\phi}&-\partial^2_{\delta r\pi}&-\partial^2_{\delta r^2}\\
    \partial^2_{\pi\theta}&\partial^2_{\pi\phi}&\partial^2_{\pi^2}&\partial^2_{\pi\delta r}
    \end{pmatrix}H
\end{equation}
so that the equation of motions can be written as 
\begin{equation}
  \dot{\bm{\xi}}=\mathcal{H}\bm{\xi}
\end{equation}
where every equation regarding the trigonometric variables $\theta$ and $\phi$ has a factor $\sin\theta$ that is adsorbed in the definition of the Hamiltonian. In Fourier space the equations of motion are equivalent to an eigenvalue problem. 
\begin{equation}
    \mathcal{H}\bm{\xi}=i\omega\bm{\xi}.
\end{equation}
By computing the electric and magnetic eigenmodes $\bar\omega^{e/m}_i$ we are able to express the electric or magnetic susceptibilities, as shown in the main text, as 
\begin{equation}
    \alpha^{e/m}(\omega)=\sum_{i}\frac{K^{e/m}}{(\bar\omega^{e/m}_i)^2-\omega^2}  
\end{equation}
where $K^{e/m}$ is a dimensional constant. 

{\large \textbf{S2. Role of DW and DM interaction}}\\
The following figures illustrate different contributions to the response functions and the effect of domain walls and of DM interactions on the susceptibilities. Note different susceptibility ranges on the colorscales.

\begin{figure}[ht]
    \centering
    \includegraphics[width=\lw]{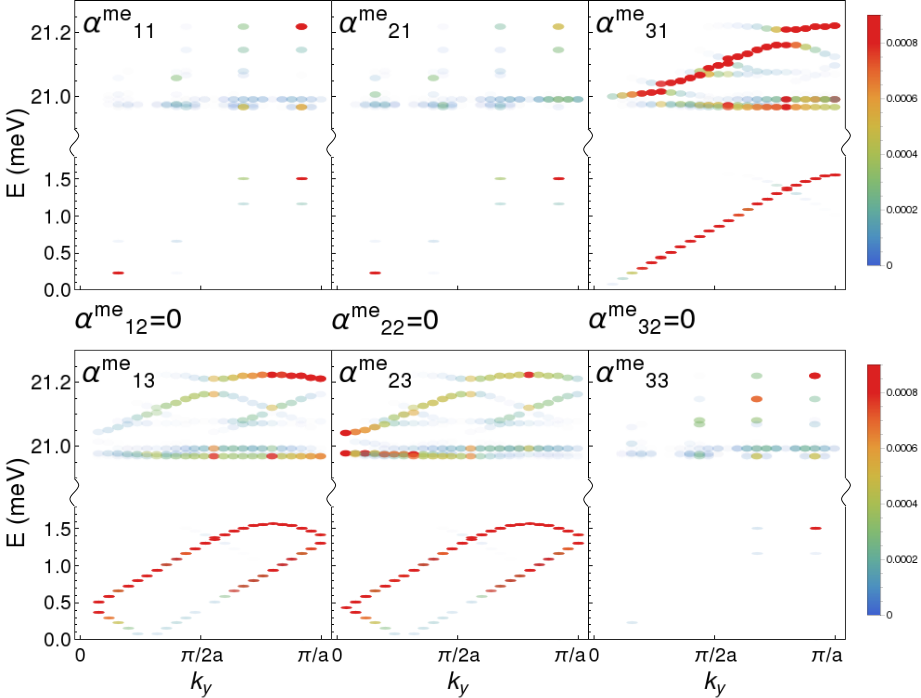}
    \caption{\label{supME}Components of the ME susceptibility computed with no DW and DM interaction turned on}
\end{figure}

\begin{figure}
    \centering
    \includegraphics[width=\lw]{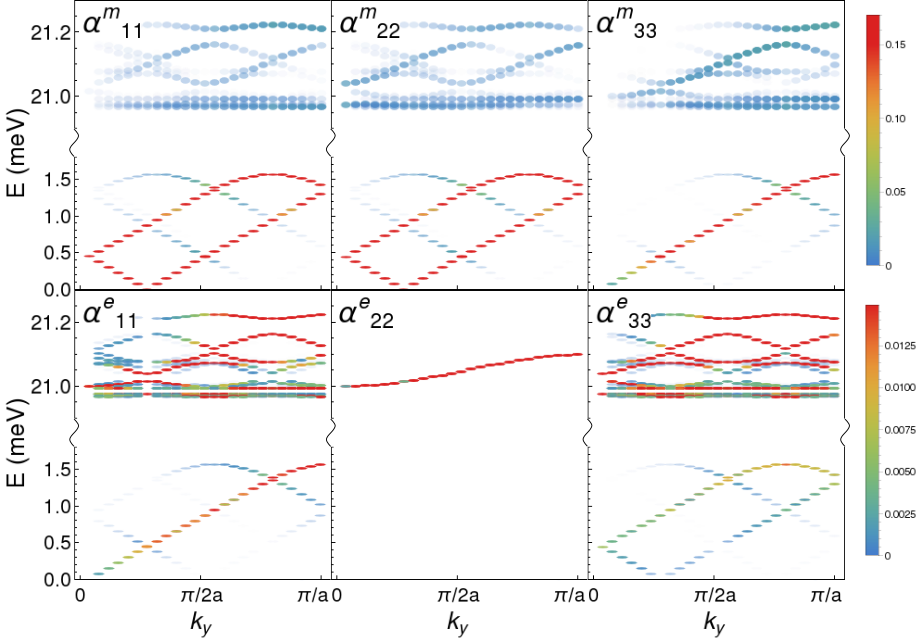}
    \caption{Components of the ME susceptibility computed with no DW, as in Fig.~\ref{supME}, but with DM interactions included}
    \label{supX1}
\end{figure}

\begin{figure}
    \centering
    \includegraphics[width=\lw]{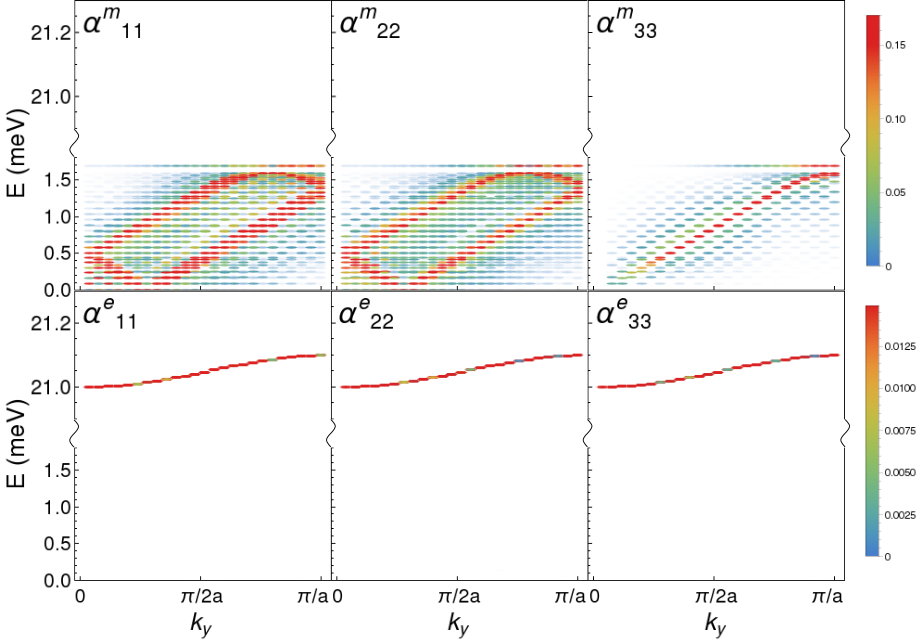}
    \caption{Components of the ME susceptibility computed with the DW present, but with DM interaction switched off (hence the empty high energy part for the magnetic susceptibility and low energy response in the electric one). The DW affects the magnons but the polar mode is unaffected due to the lack of DM interactions. It is interesting to see that a slightly gapped mode is present at 1.6 meV and this is not a boundary effect.}
    \label{supX2}
\end{figure}

\begin{figure}
    \centering
    \includegraphics[width=\lw]{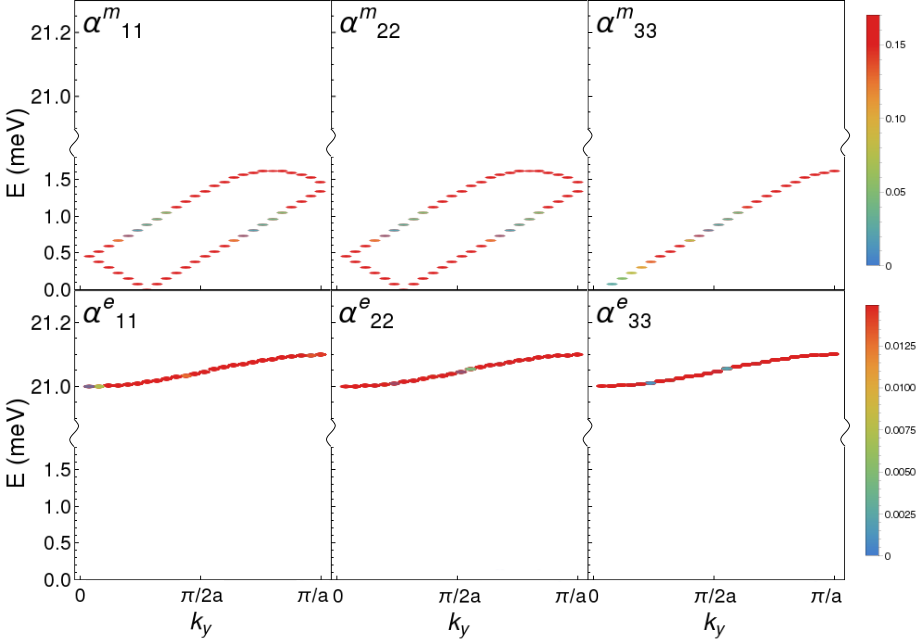}
    \caption{Components of the ME susceptibility computed without domain walls and with the DM interaction turned off. We just have the spiral magnon with the folding of the Brillouin zone but the absence of DW means that the band does not get wider due to magnon scattering.}
    \label{supX3}
\end{figure}

\begin{figure}
    \centering
    \includegraphics[width=\lw]{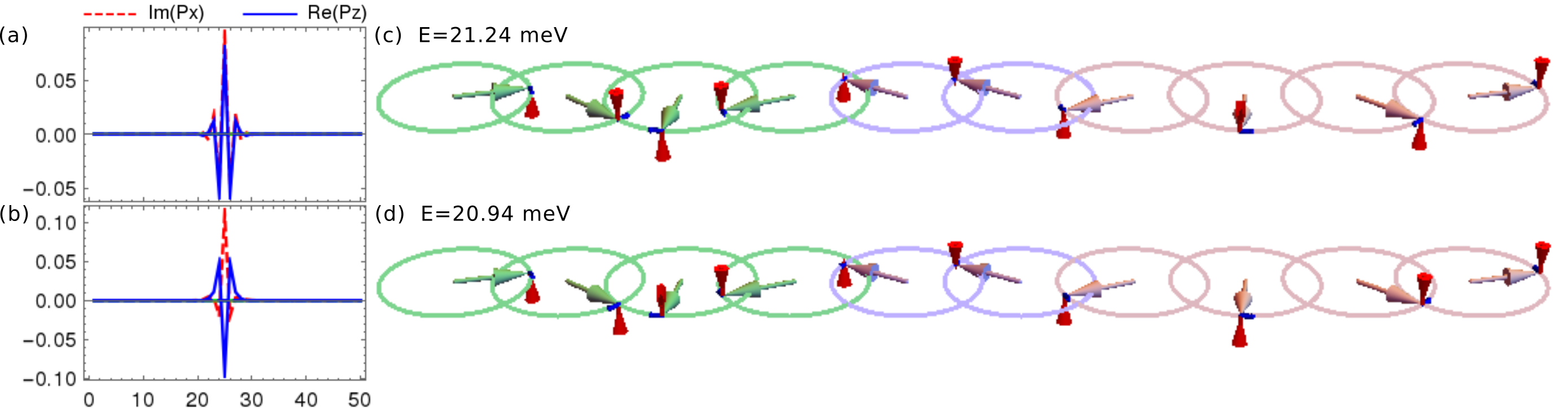}
    \caption{Dispersionless modes. In the high energy part of the spectrum two dispersionless modes appear at the highest and lowest ends of that energy interval. They are both localized modes.}
    \label{fig:my_label}
\end{figure}

\end{document}